\documentclass{llncs}
\usepackage{graphicx}

\begin{document}

\title{Analyzing the Facebook Friendship Graph}
\titlerunning{Analyzing the Facebook Friendship Graph} 
\author{Salvatore Catanese\inst{1}, Pasquale De Meo\inst{2}, Emilio Ferrara\inst{3}, Giacomo Fiumara\inst{1}}
\authorrunning{Catanese et al.} 
%
%
\institute{Dept. of Physics, Informatics Section. University of Messina.\\
\email{salvocatanese@gmail.com; giacomo.fiumara@unime.it}
\and
Dept. of Computer Sciences, Vrije Universiteit Amsterdam.\\
\email{pasquale.de.meo@few.vu.nl}
\and
Dept. of Mathematics. University of Messina.\\
\email{emilio.ferrara@unime.it}
}

\maketitle              

\begin{abstract}
Online Social Networks (OSN) during last years acquired a huge and increasing popularity as one of the most important emerging Web phenomena, deeply modifying the behavior of users and contributing to build a solid substrate of connections and relationships among people using the Web. In this preliminary work paper, our purpose is to analyze Facebook, considering a significant sample of data reflecting relationships among subscribed users. Our goal is to extract, from this platform, relevant information about the distribution of these relations and exploit tools and algorithms provided by the Social Network Analysis (SNA) to discover and, possibly, understand underlying similarities between the developing of OSN and real-life social networks. 


\end{abstract}
\section{Introduction}
The problem of analyzing social networks was already introduced during late sixties by Milgram \cite{Milgram1967} and Travers \cite{Travers1969} in Psychology and Sociology. 
Starting from this point for twenty and more years, several kind of real-life social experiments have been conducted and studied by sociologists, trying to understand motivations, dynamics and rules of real-life social networks. 

During last years the Web phenomenon of OSN started spreading and computational aspects have also been considered \cite{barabasi,kleinberg,golbeck}.
Several Social Networking Services were developed, most of them gathered millions of users in an incredible short amount of time. 
The OSN we are considering in this work is Facebook \footnote{http://www.facebook.com}, which collected more than 500 millions of world-wide users as of July 2010.

The unpredicted success and the fast growing rate of these platforms shortly opened new fascinating academic problems; e.g. it is possible to study OSN with tools provided by the SNA science \cite{Mislove2007}? 
Is the behavior of OSN users comparable with the one showed by actors of real-life social networks \cite{garton97}? 
What are the topological characteristics of OSN \cite{Ahn2007}? 
And what about their structure and evolution\cite{Kumar2006}? 
Today we exploit computational resources to analyze data acquired from OSN, trying to answer to these problems. 
In this work, we analyze connections among almost a million of Facebook users, data collected through some developed \emph{ad hoc} Information Extraction techniques.

This paper is organized as follows: in Section 2 we consider related works on social networks, OSN, etc., in particular regarding data mining experiments and SNA; Section 3 covers aspects of Artificial Intelligence and Information Extraction related to algorithms and techniques used in order to acquire and gather data from Facebook; Section 4 presents collected data, focusing on their statistical analysis, exploiting some tools provided by the SNA science. 
In Section 5 we try to graphically plot this information, i.e., a large graph where nodes represent users and edges reflect ties among them. 
Section 6 concludes, providing some suggestions for future work.

\section{Related Work}
Literature on Web (and social Web) data extraction is growing: Ferrara et al. \cite{Baumgartner2010} provided a comprehensive survey on applications and techniques. 
In \cite{Ferrara2010}, Ferrara and Baumgartner developed some techniques for automatic wrapper adaptation. 
A slightly modified version of that algorithm, relying on analyzing structural similarities inside the DOM tree structure of Facebook friend-list pages, is the core of the agent used here to gather data.

A common SNA task is to discover, if existing, aggregations and subsets of nodes playing similar roles or occupying a particular position in a network \cite{Carrington2005}. 
Some strictly connected problems are related to optimizing the visual representation of graphs \cite{Battista1994}; for large social networks graphs is not trivial to find a meaningful graphical representation, because of the number of elements to display, and finding algorithms for the planar embedding of the graph, so as reducing (or eliminating) intersecting edges and improving aesthetic and functional characteristics of the graph itself, is part of the solution \cite{Boyer2004}.

Several SNA tools have been developed during the last years: GUESS \cite{Adar2006} focuses on improving the interactive exploration of graphs; NodeXL \cite{Smith2009}, developed as an add-in to the Microsoft Excel 2007 spreadsheet software, provides tools for network overview, discovery and exploration. 
LogAnalysis \cite{Catanese2010} helps forensic analysts in visual statistical analysis of mobile phone traffic networks. 
Jung \cite{Madadhain2005} and Prefuse \cite{Heer2005} provide Java APIs implementing algorithms and methods for building applications for graphical visualization and SNA for graphs.

\section{Data Extraction and Data Cleaning Aspects}
The very first step of a SNA experiment is acquiring data: for this purpose we designed and developed a custom agent, an automaton simulating the behavior of real users, visiting Facebook publicly accessible profiles and automatically extracting relationships among them. 
Once acquired, information must be collected in some kind of well-structured format; completed this process, data must be cleaned, removing duplicates and irrelevant information, then they are ready to be used for their purpose.
 

In order to acquire information about friendship relations, we developed an agent that automatically visits the friend-list page of a real user seed profile, and then recursively, acquires friendship relations visiting friend-list pages of friends of the seed, and so on, down to the third sub-level of friendship relations. 
Only friendship relations among real users have been acquired, fan pages and companies having been discarded (Facebook provides this filter).
This agent acquires information only from profiles in friendship relation with the seed and from publicly accessible profiles, thus respecting the Facebook privacy policies.

We thus obtained an undirected graph composed of 547,302 vertices and 836,468 edges; for privacy reasons only user IDs were collected. 
The agent was developed in Java, and it embeds a Firefox browser interfaced through XPCOM \footnote{https://developer.mozilla.org/en/xpcom} and XULRunner \footnote{https://developer.mozilla.org/en/XULRunner}.

Facebook profiles are saved as GraphML \cite{Brandes2002} nodes with one attribute, namely the Facebook ID. 
Friendship relations are saved as undirected edges connecting two nodes.
Because of the intrinsic nature of the data mining process, it could happen to save parallel edges and multiple instances of the same node. 
We developed a fast algorithm of data cleaning, running in $O(n \log n)$, exploiting the hash property of the Java HashSet, which, first of all removes all duplicate nodes and, then, fixes all edges in order to link the unique instance of source and target nodes and, finally, deletes the parallel ones.

\section{Data Analysis}
SNA provides some useful techniques to analyze dynamics of relationships: it is possible to identify models and flows, in the structure of the network, e.g. trying to understand which role actors play in the environment they are placed in.
Several statistical algorithms are helpful in discovering key nodes, creating groups of social cohesion \cite{Wasserman1994}. 
However, it is not trivial to discover models or anomalies while dealing with huge amount of data; in these cases, visualization of information, on the one hand, could be useful to simplify the work of analysis, but on the other hand, could be tricky for several reasons. 
The computational cost is very high and increases with the dimension of data; it becomes harder and harder to understand graphs showing thousands of nodes and edges, also because of overlapping elements. 
For these reasons we shall analyze data using filters and clustering methods.

\subsection{Metrics and Measures}
The following measures for SNA have been standardized by Perer and Shneiderman \cite{Perer2006}: overall network metrics (number of nodes, number of edges, density, diameter), node rankings (degree, betweenness and closeness centrality), edge rankings (weight, betweenness centrality), edge rankings in pairs and cohesive subgroups. A short summary of some metrics, evaluated using NodeXL, follows.

\begin{table}%
\begin{center}
\begin{tabular}{lr}
\hline
	Graph Type: & Undirected \\
		
	Vertices: & 547,302 \\
		
	Unique Edges: & 836,468 \\
	
	Edges With Duplicates: & 0 \\
	
	Total Edges: & 836,468 \\
		
	Self-Loops: & 0 \\
		
	Connected Components: & 2 \\
	
	Single-Vertex Connected Components: & 0 \\
	
	Maximum Vertices in a Connected Component: & 546,733 \\
	
	Maximum Edges in a Connected Component: &	835,951 \\
		
	Maximum Geodesic Distance (Diameter): &	10 \\
	
	Average Geodesic Distance:& 5.00\\
\hline
\end{tabular}
\caption{Overall Network Metrics}
\label{tab1}
\end{center}
\end{table}

\vspace{-1cm}
\begin{table}%
\begin{center}
\begin{tabular}{ l r r r r }
	\cline{2-5}
	\multicolumn{1}{r}{} & Minimum & Maximum & Average & Median\\
	\hline
	Degree & 1 & 4,958 & 3.057 & 1.000 \\
	
	PageRank & 0.269 & 2120,268 & 1.000 & 0.491 \\
	
	Clustering Coefficient & 0.000 & 1.000 & 0.053 & 0.000 \\
	
	Eigenvector Centrality & 0.000 & 0.003 & 0.000 & 0.000 \\
	\hline
\end{tabular}
\caption{Miscellaneous Metrics}
\label{tab2}
\end{center}
\end{table}

\vspace{-1cm}
\section{Visualization}
Analyzing large graphs is not a trivial problem: the computational cost of visualization algorithms, e.g. Fruchterman-Reingold \cite{fruchterman:graph}, Harel-Koren \cite{harelkoren}, etc., is critical, and finding useful information is hard. 
For this reason the SNA relies on filtering data calculating metrics and displaying only relevant information.

\begin{figure}[h]
	\centering
		\includegraphics[width=320pt]{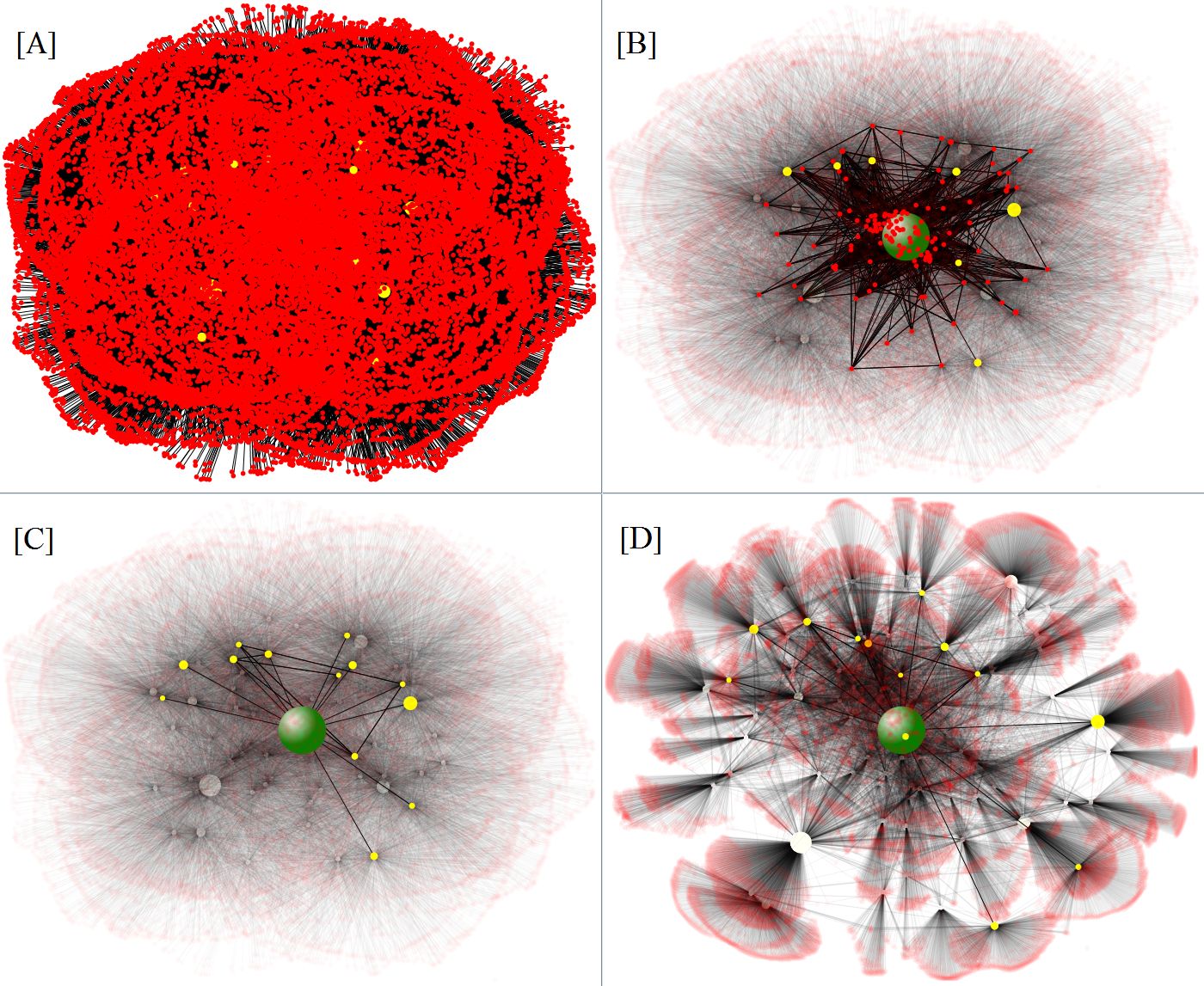}
	
	\caption{[A] Visualization of a 25,000 nodes subgraph; [B] Top 50 nodes ordered by betweenness centrality; [C] Nodes with high betweenness centrality (greater than 10 millions); [D] Clusterization after 10 iterations of the Fruchterman-Reingold algorithm.}
\label{fig:graphMashup}
\end{figure}

We produced several subgraphs for SNA and visualization purposes (see Figure \ref{fig:graphMashup}).
In Figure \ref{fig:graphMashup}B nodes are arranged according to betweenness centrality (BC). 
As the definition of BC implies, nodes in the central region of the plot show higher values of BC, thus occurring in a correspondingly higher number of shortest paths connecting all the nodes each others.
Intuitively, nodes with higher BC values have higher relevance as they represent a potential efficient way of establishing friendship relations among peripheral nodes.

\section{Conclusions}
In this preliminary work we focused on the possibility of extracting relevant information about relationships from Facebook. We developed an automaton for data mining and cleaning, gathered a sample dataset of almost a million of connections, and finally analyzed data applying SNA tools and techniques. 
Our purpose is to continue acquiring data and we already developed a more efficient way of data mining, and then to improve algorithms for data analysis and visualization, e.g. exploiting the auto parallelization and High Performance Computing techniques to handle in the most efficient way the huge amount of useful information we can gather from Facebook.
%
%
\bibliographystyle{splncs03}
\bibliography{biblio}
\end{document}